\documentclass[12pt,preprint]{emulateapj}
\usepackage{graphicx}

\newcommand{\kms}{\ensuremath{\rm km\,s^{-1}}}

\shorttitle{Detecting the Companion and Ellipsoidal Variations  of $\sigma$~Geminorum}
\shortauthors{Roettenbacher et al.}
\slugcomment{Accepted for publication in the Astrophysical Journal}

\begin{document}

\title{Detecting the Companions and Ellipsoidal Variations of RS CVn Primaries:  \\ I.  $\sigma$~Geminorum}
\author{
  Rachael M.\ Roettenbacher$^1$, John D.\ Monnier$^1$, Gregory W.\ Henry$^2$, Francis C.\ Fekel$^2$, Michael~H.~Williamson$^2$, Dimitri Pourbaix$^3$, David W.\ Latham$^4$, Christian A.\ Latham$^4$, Guillermo Torres$^4$, Fabien Baron$^1$, Xiao Che$^1$,  Stefan Kraus$^5$, Gail~H.~Schaefer$^6$, Alicia N.\ Aarnio$^1$, Heidi Korhonen$^7$, Robert~O.~Harmon$^8$, Theo A.\ ten Brummelaar$^6$, Judit~Sturmann$^6$, Laszlo Sturmann$^6$, and Nils H.\ Turner$^6$
  }
  \affil{$^1$Department of Astronomy, University of Michigan, Ann Arbor, MI 48109, USA \\  
  $^2$Center of Excellence in Information Systems, Tennessee State University, Nashville, TN 37209, USA \\ 
 $^3$FNRS, Institut d'Astronomie et d'Astrophysique, Universit\'e Libre de Bruxelles (ULB), Belgium \\
 $^4$Harvard-Smithsonian Center for Astrophysics, 60 Garden Street, Cambridge, MA 02138, USA \\
 $^5$School of Physics, University of Exeter, Stocker Road, Exeter, EX4 4QL, UK\\
  $^6$Center for High Angular Resolution Astronomy, Georgia State University, Mount Wilson, CA 91023, USA \\
 $^7$Finnish Centre for Astronomy with ESO (FINCA), University of Turku, V\"ais\"al\"antie 20, FI-21500 Piikki\"o, Finland \\
 $^8$Department of Physics and Astronomy, Ohio Wesleyan University, Delaware, OH 43015, USA 
   }
\email{rmroett@umich.edu}
%%%%%%%%%%%%%%%%%%%%%%%%%%%%%%%%%%%%%%%%%%%%%%%%%%%%%%%%%%%%%%%%%%%%

\begin{abstract}

To measure the properties of both components of the RS CVn binary $\sigma$~Geminorum ($\sigma$~Gem), we directly detect the faint companion, measure the orbit, obtain model-independent masses and evolutionary histories, detect ellipsoidal variations of the primary caused by the gravity of the companion, and measure gravity darkening.   We detect the companion with interferometric observations obtained with the Michigan InfraRed Combiner (MIRC) at Georgia State University's Center for High Angular Resolution Astronomy (CHARA) Array with a primary-to-secondary $H$-band flux ratio of $270\pm70$.   A radial velocity curve of the companion was obtained with spectra from the Tillinghast Reflector Echelle Spectrograph (TRES) on the 1.5-m Tillinghast Reflector at Fred Lawrence Whipple Observatory (FLWO).  We additionally use new observations from the Tennessee State University Automated Spectroscopic and Photometric Telescopes (AST and APT, respectively).  From our orbit, we determine model-independent masses of the components ($M_1=1.28\pm0.07$~$M_\odot$, $M_2=0.73\pm0.03$~$M_\odot$), and estimate a system age of $5\mp1$~Gyr.   An average of the $27$-year APT light curve of $\sigma$~Gem folded over the orbital period ($P=19.6027\pm0.0005$~days) reveals a quasi-sinusoidal signature, which has previously been attributed to active longitudes $180^\circ$ apart on the surface of $\sigma$~Gem.  With the component masses, diameters, and orbit, we find that the predicted light curve for ellipsoidal variations due to the primary star partially filling its Roche lobe potential matches well with the observed average light curve, offering a compelling alternative explanation to the active longitudes hypothesis.  Measuring gravity darkening from the light curve gives $\beta < 0.1$, a value slightly lower than that expected from recent theory.

\end{abstract}


%%%%%%%%%%%%%%%%%%%%%%%%%%%%%%%%%%%%%%%%%%%%%%%%%%%%%%%%%%%%%%%%%%%%

\section{Introduction}

RS~Canum~Venaticorum (RS~CVn) stars are spotted, active binary systems exhibiting photometric and Ca~H~and~K variability  \citep{hal76}.  Often tidally-locked, these systems are composed of an evolved primary star (giant or subgiant) and a subgiant or dwarf companion \citep{ber05,str09}.  With active binaries, not only is there potential to determine the component masses and system evolutionary history but also to understand the magnetic field interactions through active longitudes, particular longitudes $180^\circ$ apart with persistent, long-lived starspots \citep{ber98,ber05}.  

Observing the magnetic phenomena of rapidly-rotating evolved stars also sheds light on the magnetic activity of rapidly-rotating young stars, such as T~Tauri stars.  Both T~Tauri and RS~CVn systems have starspots analogous to sunspots---cool starspots resulting from stifled convection in the outer layers of the stars due to strong magnetic fields \citep{pet03, ber05}.

$\sigma$~Geminorum ($\sigma$~Gem, HD~62044, HIP~37629, HR~2973) is an RS~CVn system known to exhibit starspots, often ascribed to ``active longitudes'' \citep[e.g.,][]{hal77, hen95}.  The system has been characterized as a single-lined spectroscopic binary \citep{her55} with a K1III primary \citep{rom52}.  The orbital period of $\sigma$~Gem is slightly longer than the primary star's rotation period derived from the fastest rotating spots \citep[$P_\mathrm{orb} = 19.60$ days, $P_\mathrm{rot, min} = 19.47$ days;][]{kaj14}.    

Because of its large starspots, $\sigma$~Gem is a frequent target for understanding starspot evolution.  \citet{ebe13} first reported $\sigma$~Gem as active and potentially spotted due to fluctuations in the Ca~H~and~K lines as the star rotated.  Decades later, \citet{hal77} identified photometric variations suggesting starspots ($\Delta V \sim 0.07$).  Initial models of the surface of $\sigma$~Gem often showed the surface with two starspots oriented on opposite sides of the primary star \citep{fri83}.  \citet{ber98} emphasize that, due to tidal locking, the starspots are located such that one spot constantly faces the companion and the other spot is $180^\circ$ offset.   The majority of spot models applied to light curves of $\sigma$~Gem consist of two spots on a spherical star \citep{eke86, str88, ola89, hen95, jet96, pad99, kaj14}.  Doppler images have suggested the surface is covered with a larger number of smaller spots \citep{hat93,kov01,kov14}. 

To understand the binary system, we present our analysis of the first detections of the companion in our interferometric and radial velocity data sets, as well as photometric data.  In Section 2, we describe the observations for our data sets.  In Section 3, we discuss our analysis of the data sets, including the first astrometric and spectroscopic detections of the companion star and orbital parameters.  In Section 4, we present evolutionary constraints and a Hertzsprung-Russell (H-R) diagram.  In Section 5, we discuss our analysis of the photometric data set, including detected ellipsoidal variations and measured gravity darkening.  In Section 6, we present the conclusions of our study of $\sigma$~Gem.

%%%%%%%%%%%%%%%%%%%%%%%%%%%%%%%%%%%%%%%%%%%%%%%%%%%%%%%%%%%%%%%%%%%%

\section{Observations}

\subsection{Interferometry}

We obtained interferometric data with Georgia State University's Center for High-Angular Resolution Astronomy (CHARA) Array.  The CHARA Array is a Y-shaped array of six $1$-m class telescopes with non-redundant baselines varying from $34$- to $331$-m located at Mount Wilson Observatory, California \citep{ten05}.  Using all six telescopes and the Michigan InfraRed Combiner \citep[MIRC;][]{mon04, mon06}, we obtained $H$-band ($1.5-1.8 \ \mu$m) data (eight channels across the photometric band with $\lambda/\Delta \lambda \sim 40$) on UT 2011 Nov 9 and Dec 7, 8, 9; 2012 Nov 7, 8, 21, 22, 24, 25 and Dec 4, 5.  

We made detections of the companion in the data from UT 2011 Dec 8; 2012 Nov 7, 8, 24, and 25.  The remaining nights of observation had insufficient $uv$ coverage due to poor seeing or short observation lengths, leaving the companion undetected.  We reduced and calibrated these data with the standard MIRC pipeline \citep[see][for pipeline details]{mon07,mon12,zha09,che11}.  We used at least one calibration star for each night of data (see Table \ref{cal}).  

\begin{deluxetable*}{l c c c}
\tabletypesize{\scriptsize}
\tablecaption{Calibrators for $\sigma$~Geminorum}
\tablewidth{0pt}
\tablehead{
\colhead{Calibrator Name} & \colhead{Calibrator Size (mas)} & \colhead{Source} & \colhead{UT Date of Observation}
}
\startdata
HD 37329 & $0.71 \pm   0.05$ & \cite{bon06} & 2012 Nov 8\\
HD 50019 ($\theta$ Gem) &  $0.81 \pm  0.06$ & \cite{bon06} &  2012 Nov 7, 8, 25 \\
HD 63138 & $0.65 \pm  0.04$ & MIRC calibration & 2011 Dec 8; 2012 Nov 8\\
HD 69897 ($\chi$ Cnc) & $0.73 \pm 0.05$ & \cite{bon06} & 2012 Nov 7, 24, 25
\enddata
\label{cal}
\end{deluxetable*}

\subsection{Radial Velocity}

To constrain the spectroscopic orbit for $\sigma$~Gem, we utilized
three independent sets of radial-velocity data: two sets of
single-lined velocities for the primary, and a new set of double-lined
velocities for both components of the binary.

One set of the radial velocity measurements for the primary star was published in \citet{mas08}.  These 39 data points were obtained with two identical CfA Digital Speedometers \citep{lat92} on the $1.5$-m Wyeth Reflector (Oak Ridge Observatory) and $1.5$-m Tillinghast Reflector (Fred Lawrence Whipple Observatory) telescopes 
(2003 December 30 $-$ 2007 June 5). 

From 2012 October 1 $-$ 2015 January 9, using the Tillinghast telescope with the Tillinghast Reflector Echelle Spectrograph \citep[TRES; ][]{fur08}, we were able to make fifteen detections of the secondary spectra for the first time.  Along with sixteen new primary star measurments, these new radial velocities are presented in Table \ref{RVCfA}.
 We add 0.14 \kms to these sets of radial velocities to account for these data being reported on the CfA native system \citep[][note the correction is inaccurately stated as a subtraction in this reference]{ste99}. For details of these observations and data analysis, see Appendix A.

\begin{deluxetable}{l c c}
\tabletypesize{\scriptsize}
\tablecaption{Radial Velocity Data of $\sigma$~Gem (CfA)}
\tablewidth{0pt}
\tablehead{
\colhead{HJD $-2400000$} & \colhead{Primary (\kms)} & \colhead{Secondary (\kms)} 
}
\startdata
    56202.0199  & 10.66 & 103.44\\
    56230.0436  & 77.98 &  -9.80\\
    57002.9567  & 14.17 &  94.68\\
    57003.9301  &  9.71  & 101.50\\
    57014.8600  & 79.02 & -12.11\\ 
    57015.8427  & 75.96 & -10.77\\
    57018.8519  & 51.08 &  28.91\\
    57019.9181  & 38.76 &  60.57\\ 
    57020.9452  & 27.84 &  78.02\\
    57021.8962  & 19.05 &  84.20\\
    57024.9812  &  9.10  & 104.10\\
    57025.9317  & 12.59 &  99.41\\
    57026.8891  & 18.88 &  85.69\\
    57028.9057  & 38.58 & \\
    57029.8670  & 49.81 &  27.28\\
    57031.8806  & 69.76 &  -2.88
\enddata
\tablecomments{Errors on the primary radial velocities are $0.84$ \kms.  Errors on the secondary radial velocities are $3.8$ \kms.  These were then scaled for our orbit fit to have a total $\chi^2 = 1.00$.  Note these radial velocities are on the native CfA system.  We added $0.14$ \kms for use in our analysis \citep{ste99}.} 
\label{RVCfA}
\end{deluxetable}

The additional radial velocity data set consists of 43 spectrograms of the primary star of
$\sigma$~Gem taken between 
2009 January 12 $-$ 2014 December 1 with the Tennessee State
University 2-m automatic spectroscopic telescope (AST), fiber-fed
echelle spectrograph, and a CCD detector at Fairborn Observatory, Arizona \citep[see Table \ref{RVfekel};][]{eat04,eat07}. At first, the detector
was a 2048 $\times$ 4096 SITe ST-002A CCD with 15~$\mu$m pixels.
\citet{eat07} discussed the reduction of the raw spectra and
wavelength calibration. Those echelle spectrograms have 21 orders
that cover the wavelength range 4920--7100~\AA\ with an average
resolution of 0.17~\AA, corresponding to a resolving power of
35000 at 6000~\AA. Those spectra have a typical signal-to-noise
value of 30.

In the summer of 2011 the AST SITe CCD and its dewar were retired and replaced with a Fairchild 486 CCD, a 4096 $\times$ 4096 array of 15~$\mu$m pixel, that is housed in a new dewar. With the new CCD the wavelength coverage ranged from 3800 to 8600~\AA. The resolution was reduced slightly to 0.24~\AA\ or a resolving power of 25000 at 6000~\AA. These more recent spectra have signal-to-noise ratios of about 70.

\citet{fek09} provided an extensive general description of velocity
measurement of the Fairborn AST spectra.  In the case of $\sigma$~Gem,
we measured a subset of 63 lines from our solar-type star line
list that covers the 4920--7120~\AA\ region. Because the lines of
$\sigma$~Gem have significant rotational broadening, we fit
the individual lines with a rotational broadening function. The
Fairborn velocities are on an absolute scale. A comparison of our
unpublished measurements of several IAU standard stars with those
determined by \citet{sca90} indicates that the Fairborn Observatory
velocities from the SITe CCD have a small zero-point offset of
$-$0.3 \kms.  Velocities from the Fairchild CCD spectra
have a slightly larger zero-point offset of $-$0.6 \kms
relative to those of \citet{sca90}. 
Thus, in Table \ref{RVfekel} we corrected our measured
velocities by either 0.3 or 0.6 \kms, depending on which
detector was used.

\begin{deluxetable}{l c}
\tabletypesize{\scriptsize}
\tablecaption{Radial Velocity Data of $\sigma$~Gem (AST/TSU)}
\tablewidth{0pt}
\tablehead{
\colhead{HJD $-2400000$} & \colhead{Primary (\kms)} 
}
\startdata
       54843.9716     &        38.0     \\
       54844.8597     &        28.4     \\
       54845.6849     &        20.5     \\
       54846.6455     &        14.0     \\
       54847.6848     &        9.2     \\
       54848.7712     &        9.2     \\
       54849.7451     &        12.1     \\
       54850.6645     &        17.6     \\
       54856.9772     &        75.7     \\
       54859.7922     &        73.1     \\
       54860.6910     &        66.7     \\
       54861.6447     &        58.7     \\
       54862.6829     &        48.3     \\
       54863.6434     &        37.3     \\
       54864.6443     &        26.7     \\
       54865.6596     &        17.8     \\
       54866.6401     &        11.7     \\
       54867.6456     &        9.1     \\
       54868.9572     &        10.3     \\
       54869.6594     &        13.8     \\
       54870.6346     &        21.1     \\
       54975.6558     &        77.9     \\
       54976.6372     &        76.4     \\
       54981.6497     &        33.3     \\
       54982.6541     &        23.0     \\
       54983.6544     &        15.2     \\
       54984.6546     &        9.9     \\
       54985.6548     &        8.5     \\
       54986.6547     &        10.7     \\
       55060.9627     &        23.5     \\
       55061.9642     &        16.0     \\
       55066.9516     &        23.3     \\
       55069.9622     &        54.3     \\
       55070.9596     &        64.0     \\
       55071.9572     &        71.9     \\
       55072.9375     &        76.6     \\
       56984.7820     &        9.4     \\
       56987.7638     &        19.7     \\
       56988.7240     &        28.0     \\
       56989.7317     &        38.9     \\
       56990.7354     &        50.0     \\
       56991.7355     &        60.9     \\
       56992.8661     &        70.8   
\enddata
\tablecomments{Errors on the primary radial velocities are $0.3$ \kms.  These data were then scaled for our orbit fit to have a total $\chi^2 = 1.00$.  } 
\label{RVfekel}
\end{deluxetable}

\subsection{Photometry}

We used differential photometry of $\sigma$~Gem and a comparison star from the Tennessee State University T3 0.4-m Automated Photometric Telescope (APT) located at Fairborn Observatory, Arizona.  For details on the observational procedure and photometers see \citet{hen99} and \citet{fek05}.

\begin{figure}
\hspace{-0.3cm}
\includegraphics[angle=90,scale=0.38]{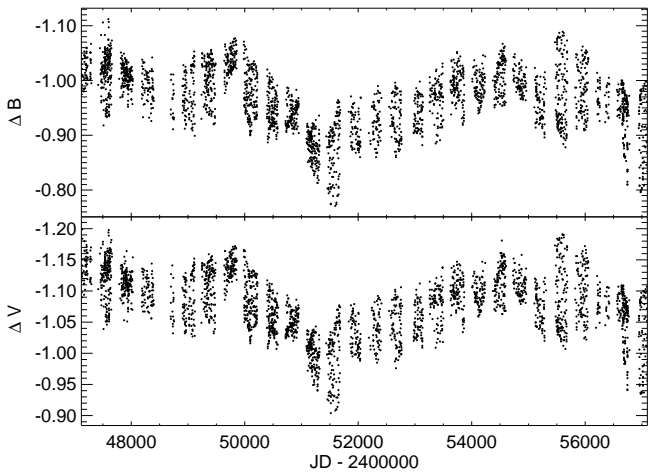}
\caption{Johnson $B$ and $V$ differential magnitudes of $\sigma$~Gem acquired over 28 observing seasons from 1987 $-$ 2015 with the T3 $0.4$-meter APT at Fairborn Observatory in southern Arizona.}
\label{fullLC}
\end{figure}

The differential Johnson \emph{B} and \emph{V} light curves cover 
1987 November 21 $-$ 2015 March 13 (see Table \ref{LC} and Figure \ref{fullLC}).  Subsets of these data were analyzed by \citet{hen95} and \citet{kaj14}.  For the first time, we make the full set of T3 APT photometry of $\sigma$~Gem available in electronic format in Table \ref{LC}.

% stub version
\begin{deluxetable*}{l c c c c}
\tabletypesize{\scriptsize}
\tablecaption{Johnson $B$ and $V$ Differential Magnitudes of $\sigma$~Gem from the T3 APT}
\tablewidth{0pt}
\tablehead{        
\colhead{HJD $-2400000$} &  \colhead{$\Delta B$ ($\sigma$~Gem - HD~60318$^a$)} & \colhead{$\Delta V$ ($\sigma$~Gem - HD~60318)} & \colhead{$\Delta B$ ($\upsilon$~Gem$^b$ - HD~60318)} & \colhead{$\Delta V$ ($\upsilon$~Gem - HD~60318)}
}
\startdata
47121.0481		&	-1.006		&	-1.123		&	     		&	     		\\
47122.0465		&	-1.010		&	-1.135		&	     		&	     		\\
47125.0364		&	-1.030		&	-1.125		&	-0.766		&	-1.303		\\
47126.0364		&	-1.011		&	-1.128		&	-0.757		&	     		\\
47128.0310		&	-0.975		&	     		&	-0.759		&	                     
\enddata
\tablecomments{Table \ref{LC} is published in its entirety in the electronic edition.  A portion is shown here for guidance regarding its form and content.\\
$^a$  HD~60318, $V$ = 5.33, $B-V$ = 1.01 \\
$^b$  $\upsilon$~Gem (HD~60522, 69~Gem), $V$ = 4.06, $B-V$ = 1.54
} 
\label{LC}
\end{deluxetable*}

%%%%%%%%%%%%%%%%%%%%%%%%%%%%%%%%%%%%%%%%%%%%%%%%%%%%%%%%%%%%%%%%%%%%

\section{Orbital Elements}

In order to derive the astrometric orbit of $\sigma$~Gem, we searched for the companion with model fitting.  We modeled the system with the resolved primary star and an unresolved secondary.  We allowed the primary radius along the major axis, primary major-to-minor axis ratio, primary major axis position angle, primary-to-secondary flux ratio, and secondary position to vary.  During the fitting, we weighted the data such that the separate observables (squared visibilities, closure phases, and triple amplitudes) contributed to the final $\chi^2$ with equal weight.  The parameter errors for the primary star size and the primary-to-secondary flux ratio were based on the epoch-to-epoch variation, while the relative positional error of the secondary compared to the primary were based on the residuals to the orbit fit (see discussion on orbit fitting). 

The coordinates of the detections on five nights (UT 2011 Dec 8; 2012 Nov 7, 8, 24, and 25 are listed in Table \ref{detect}).  The $H$-band flux ratio for the primary star to the secondary is $270\pm70$.  In addition to detecting the secondary star, we measured the uniform disk diameter of the primary to be $\theta_{\mathrm{UD},1} = 2.335 \pm 0.007$~mas (limb-darkened disk diameter $\theta_{\mathrm{LD},1} = 2.417 \pm 0.007$~mas) with a major-to-minor axis ratio of $1.02\pm0.03$.  Our measurements are slightly larger than those in the CHARM2 catalog \citep[uniform disk diameter of $\theta_{\mathrm{UD},1} = 2.18 \pm 0.05$~mas, limb-darkened disk diameter of $\theta_{\mathrm{LD},1} = 2.31 \pm 0.05$~mas;][]{ric05}.

\begin{deluxetable*}{l c c c c c c c}
\tabletypesize{\scriptsize}
\tablecaption{Detections for the companion of $\sigma$~Geminorum with respect to the primary}
\tablewidth{0pt}
\tablehead{
\colhead{UT Date} & \colhead{JD $-2400000$} & \colhead{Separation} & \colhead{Position} & \colhead{Error Ellipse} & \colhead{Error Ellipse} & \colhead{Error Ellipse} & \colhead{Reduced} \\
 & &  \colhead{(mas)} & \colhead{Angle ($^\circ$)$^a$} & \colhead{Major Axis (mas)$^b$} & \colhead{Minor Axis (mas)$^b$} & \colhead{Position Angle ($^\circ$)$^a$} & \colhead{ $\chi^2$}
 } 
\startdata
2011 December 08 & 55903.95 & 2.83 & 19.1 & 0.30 & 0.09 & 80 & 4.4 \\
2012 November 07 & 56238.97 & 4.32  & 8.6  & 0.04 & 0.03 & 280 & 2.8 \\
2012 November 08 & 56239.86 & 4.68 & 359.7 &   0.13 & 0.06  & 300 & 2.1 \\
2012 November 24 & 56256.00 & 2.03 &  39.0 & 0.08 & 0.06 & 30 & 1.5 \\  
2012 November 25 & 56256.95 & 3.12  & 21.6 & 0.05 & 0.04 & 320 & 1.7
\enddata
\tablecomments{These detections give an $H$-band ($1.5-1.8 \ \mu$m) flux ratio for $\sigma$~Gem primary to secondary of $270\pm70$.  The uniform disk fit for the primary star is $\theta_{\mathrm{UD,1}} = 2.335\pm0.006$~mas (limb-darkened disk diameter $\theta_{\mathrm{LD,1}} = 2.417\pm0.006$~mas) with a $1.02\pm0.03$ major-to-minor axis ratio. \\
$^a$East of North\\
$^b$Scaled error bars to ensure a total $\chi^2 = 1.00$ as described in Section 3.2. } 
\label{detect}
\end{deluxetable*}

To determine the binary orbit, we simultaneously fit our interferometric and radial velocity data with Monte Carlo realizations.  The five interferometric points are as described above, and we present the scaled error bars of the major and minor axis in Table \ref{detect} to give our fit a total $\chi^2 = 1.00$.  For the radial velocity data we combine the \citet[adding $0.14$~\kms to account for the values reported on the CfA native system]{mas08}, new CfA data, and the AST data to fit simultaneously with the astrometry.  The radial velocity errors are similarly scaled ($\mathrm{rms}_\mathrm{CfA,1} = 0.84$~\kms, $\mathrm{rms}_\mathrm{AST,1} = 0.3$ \kms, $\mathrm{rms}_\mathrm{CfA,2} = 3.8$ \kms).

Using the complete radial velocity data sets, we find an eccentricity of $e = 0.014 \pm 0.004$, consistent with slightly eccentric orbits reported by \citet{har35}, \citet{pou04}, and \citet{mas08}.  However, \citet{luy36}, \citet{bat78}, and \citet{due97} reported a circular orbit.  To investigate this discrepancy, we used the APT light curve to eliminate the primary star's radial velocity data that were obtained when $\sigma$~Gem presented starspots ($\Delta V > 0.04$), as these could cause shifts in the velocities \citep[e.g.,][]{saa97}.  The remaining primary star radial velocity data obtained when $\sigma$~Gem did not exhibit large starspots from the \citet{mas08}/CfA data set span 
2006 December 6 $-$ 2007 June 5, and those from the AST data set span 
2009 January 12 $-$ Jun 4.  Using the primary star's truncated data set with $42\%$ of the \citet{mas08}/CfA and $33\%$ of the AST epochs removed, we find the orbit is consistent with a circular orbit, $e = 0.002\pm0.002$, and we adopt a circular orbit for the rest of this paper.  

\begin{figure}
\hspace{-0.5cm}
\includegraphics[angle=90,scale=0.6]{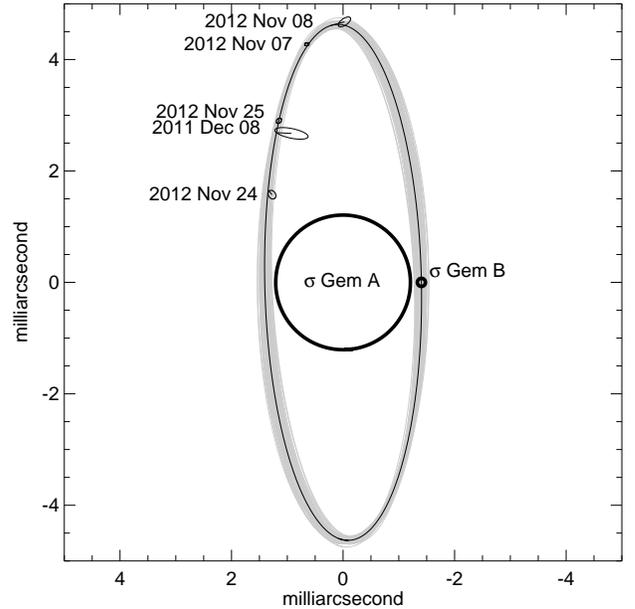}
\caption{Visual orbit for the prototypical RS~CVn system $\sigma$~Gem with our observed stellar primary radius (thick black line, $\sigma$~Gem~A) and our dates of companion detection and their locations on the orbit (black error ellipses).  The predicted radius of the companion star, $\sigma$~Gem~B, is plotted for scale with the small thick black circle.  The orbits of fifty Monte Carlo realizations are presented as the light gray orbits.  Black lines connect the center of the detection error ellipse to the expected point in the best-fit orbit, which is overlaid in black (given in Table \ref{sigGemparam} with $1-\sigma$ errors).  At the southernmost point in the orbit, the secondary star is moving toward the observer.  Note:  axis units are milliarcseconds (mas) with north upwards and east to the left.}
\label{sigGemdetection}
\end{figure}

\begin{figure}
\hspace{-0.5cm}
\vspace{0cm}
\includegraphics[angle=90,scale=.38]{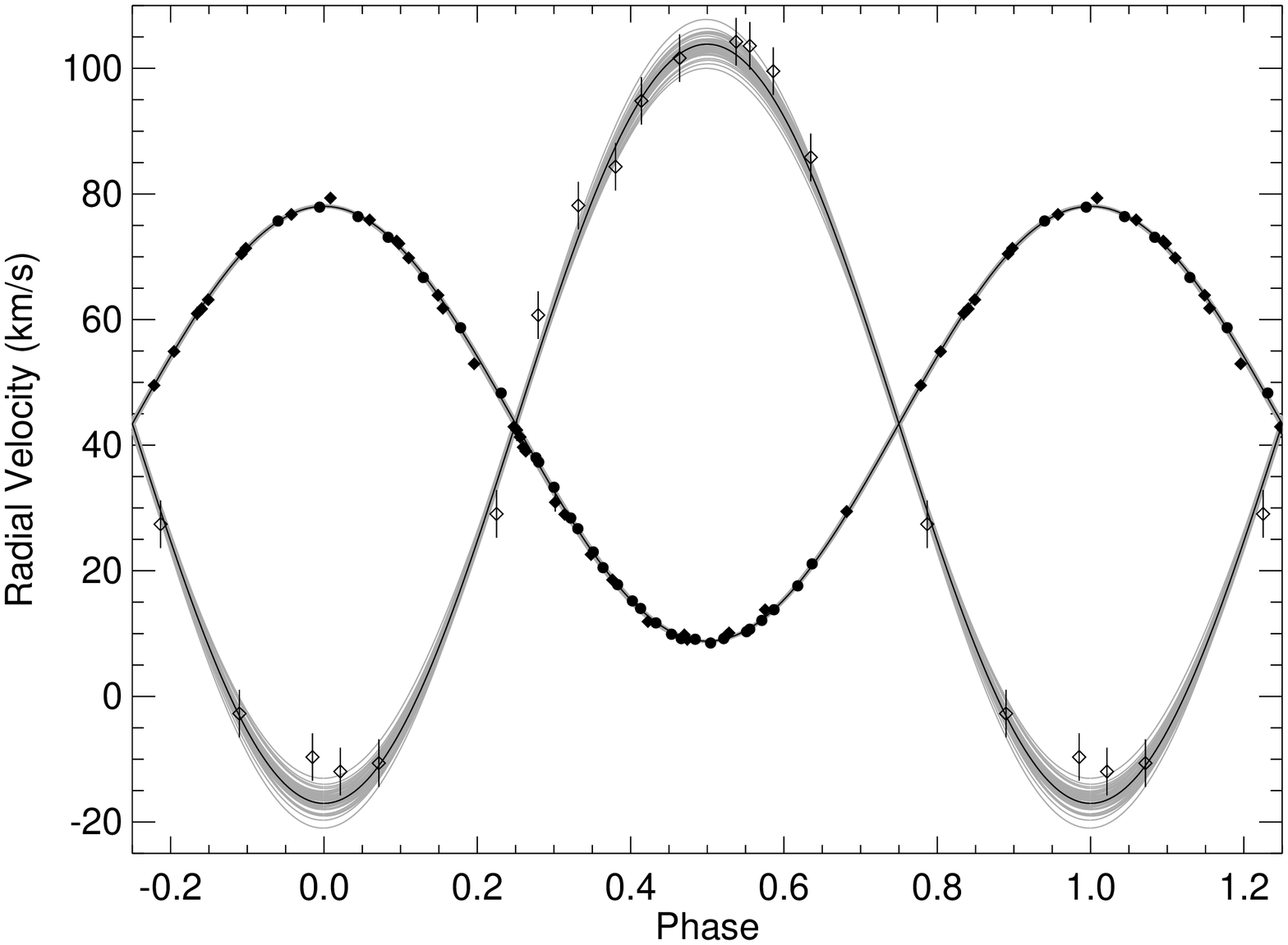}
\vspace{0cm}
\caption{Radial velocity curves of $\sigma$~Gem.  The filled diamonds represent our sample of measured observations from \citet{mas08}/CfA, and the filled circles are the AST observations.  Both data sets are restricted to those data points obtained with no starspots present (see Section 3).  $1-\sigma$ errors in velocity are presented unless the error is smaller than the diamonds and circles.  The radial velocity curves of fifty Monte Carlo realizations are presented as the light gray orbits.  The radial velocity for the best orbital parameters is overlaid in black.  Similarly, the open diamonds represent CfA radial velocity data for the secondary star with $1-\sigma$ error bars.  The light gray orbits are fifty Monte Carlo realizations with the best orbital parameters overlaid in black.  See Table \ref{sigGemparam} for orbital parameters with $1-\sigma$ errors.  }
\label{sigGemRV}
\end{figure}

Requiring eccentricity $e = 0$ and the argument of periastron for the primary $\omega = 0^\circ$ the simultaneous Monte Carlo realizations gave the orbital parameters and their $1-\sigma$ errors listed in Table \ref{sigGemparam}.  The visual orbit is illustrated in Figure \ref{sigGemdetection}, and the radial velocity curve is presented in Figure \ref{sigGemRV}.  We use the conventions presented by \citet{hei78}, where the argument of periastron, $\omega$, and the time of nodal passage (maximum recessional velocity), $T_0$, are defined by the primary star's orbit.  The ascending node, $\Omega$, is independent of definition, being equivalent with respect to either the primary or secondary star.  

Our orbital parallax, $\pi = 25.8 \pm 0.4$ mas can be compared with the Hipparcos parallax of $26.68\pm0.79$ mas \citep{ESA97}. 
As an unresolved binary with a variable component, $\sigma$~Gem does not exhibit the photocenter shifts found to be troublesome for measuring binary system parallax with Hipparcos \citep{ESA97,hal05}.   Assuming that the secondary is negligibly bright, the semi-major axis of the photocentric orbit of the primary is at most $1.71$~mas wide, which is at the limit of detectability \citep{pou02} for Hipparcos.  Combining Hipparcos data and our visual orbit, the parallax is $26.4\pm0.8$ mas, consistent with our orbital parallax.  For our subsequent analysis, we adopt our higher-precision orbital parallax, $\pi = 25.8 \pm 0.4$ mas.

With a circular orbit and $P_\mathrm{orb} \sim P_\mathrm{rot}$ \citep[e.g.,][]{kaj14}, we expect $\sigma$~Gem to have aligned rotational and orbital axes.  Given our orbital and stellar parameters, we can calculate the obliquity of the system.  Comparing our calculated value of $v \sin i = (2 \pi R_1/P_\mathrm{orb}) \times v \sin i = 24.8\pm 0.4$ \kms with the observational rotational velocity of $v \sin i = 26.7\pm0.5$ \kms (from the TRES spectra), we find that the calculation is smaller than the observational value.  This discrepancy could be attributed to the estimate of mircroturbulence or the presence of the large spot structures on the surface of $\sigma$~Gem during the TRES observations instead of a small, non-zero obliquity.  

\begin{deluxetable}{l c}
\tabletypesize{\scriptsize}
\tablecaption{Orbital and Stellar Parameters of $\sigma$~Gem}
\tablewidth{0pt}
\tablehead{
\colhead{Measured Parameters} & \colhead{Value} 
}
\startdata
semi-major axis, $a$ (mas) & $4.63\pm0.04$\\
eccentricity, $e$ & $0$\\
inclination, $i$ ($^\circ$) & $107.7\pm0.8$\\
argument of periastron, $\omega$ ($^\circ$)$^a$ & $0$\\
ascending node, $\Omega$ ($^\circ$) & $ 1.2\pm0.8$\\
period, $P_\mathrm{orb}$ (days) & $19.6027\pm0.0005$\\  
time of nodal passage, $T_0$ (HJD)$^b$ & $2453583.98\pm0.03$\\  
velocity semi-amplitude, $K_1$ (\kms) & $34.62\pm0.08$\\
velocity semi-amplitude, $K_2$ (\kms) & $60\pm2$\\
system velocity, $\gamma$ (\kms) & $43.41\pm0.08$\\
uniform disk diameter, $\theta_{\mathrm{UD,1}}$ (mas) & $2.335\pm0.007$\\  
limb-darkened disk diameter, $\theta_{\mathrm{LD},1}$ (mas)$^c$ & $2.417\pm0.007$\\  
primary major-to-minor axis ratio & $1.02 \pm 0.03$\\
$H$-band flux ratio, primary to secondary & $270 \pm 70$\\
orbital parallax, $\pi$ (mas) & $25.8\pm0.4$\\
distance, $d$ (pc) & $38.8 \pm 0.6$\\ 
\hline
\colhead{Derived Parameters} & \\
\hline
average primary radius, $R_1$ ($R_\odot$)$^d$ & $10.1\pm0.4$\\ 
primary luminosity, $L_1$ ($L_\odot$) & $39\pm2$\\  
primary surface gravity, $\log g_1$ (cm/s$^2$) & $2.54\pm0.02$\\  
primary mass, $M_1$ ($M_\odot$) & $1.28\pm0.07$\\ 
secondary mass, $M_2$ ($M_\odot$) & $0.73\pm0.03$\\
system age (Gyr) & $5 \mp1$\\
\hline
\colhead{Literature Parameters} & \\
\hline
primary effective temperature, $T_{\mathrm{eff},1}$ (K)$^e$ & $4530\pm60$\\
primary metallicity (iron), Fe/H$^f$ & 0.0
\enddata
\tablecomments{$^a$Radial velocity convention for primary with respect to the center of mass.  \\ 
$^b$Time of maximum recessional velocity of the primary star.  \\
$^c$We applied a $3.5\%$ correction from uniform to limb-darkened disk diameter.  This is equivalent to a limb-darkening coefficient $\alpha = 0.27$.\\
$^d$Using limb-darkened disk diameter.\\
$^e T_{\mathrm{eff},1}$ is an average of temperatures given by  \citet{gle79,poe85,sta94,one96,kov01,mas08}.  The $1-\sigma$ error is the standard deviation of these values. \\
$^f$[Fe/H] $= -0.02$ \citep{mal98}; approximated as [Fe/H]$ = 0.00$.} 

\label{sigGemparam}
\end{deluxetable}

%%%%%%%%%%%%%%%%%%%%%%%%%%%%%%%%%%%%%%%%%%%%%%%%%%%%%%%%%%%%%%%%%%%%

\section{Masses and Hertzsprung-Russell Diagram}

\begin{figure}
\hspace{-0.3cm}
\vspace{0cm}
\includegraphics[angle=90,scale=0.37]{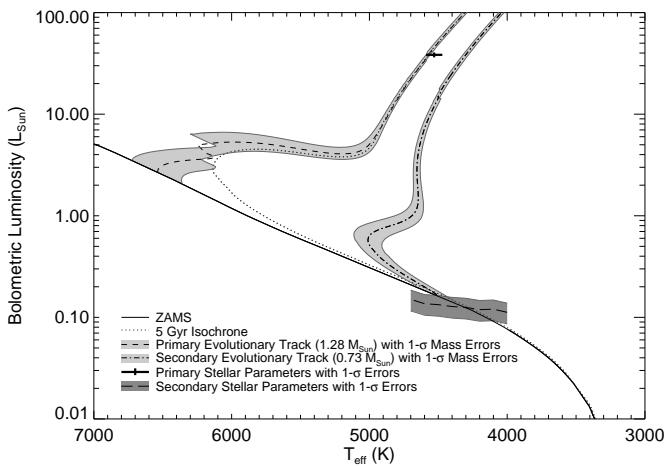}
\vspace{0cm}
\caption{H-R diagram for $\sigma$~Gem.  The dashed and dot-dashed lines are the main sequence and post-main sequence evolutionary tracks for $1.28 \ M_\odot$ and $0.73 \ M_\odot$ stars with $[Fe/H]\sim0.0$, respectively \citep{dot08}.  The gray regions represent our $1-\sigma$ mass errors ($M_{1} = 1.28 \pm 0.07 \ M_\odot; M_{2} = 0.73 \pm 0.03 \ M_\odot$) with the solid black line representing the zero age main sequence.  The dotted line is a $5$~Gyr isochrone \citep[PHOENIX;][]{dot08}.  The measured location of the primary with $1-\sigma$ errors is indicated by the plus sign.  The region where the companion could be located given our flux ratio and reasonable temperature estimates is indicated with the long-dashed line (with $1-\sigma$ errors in luminosity).}
\label{sigGemHR}
\end{figure}

Using our complete orbital fit, we obtain model-independent masses $M_1 = 1.28 \pm 0.07 \ M_\odot$ and $M_2 = 0.73 \pm 0.03 \ M_\odot$.  
With the stellar parameters of the primary star (including $T_\mathrm{eff,1} = 4530 \pm 60$~K, see Table \ref{sigGemparam}) and the primary-to-secondary $H$-band flux ratio detected using the CHARA/MIRC data ($270\pm70$), we are able to constrain the parameters (luminosity, temperature, and radius) of the secondary star.   We use the flux ratio and NextGen stellar atmospheres \citep{hau99} to constrain the stellar flux to calculate a range of luminosities for reasonable effective temperatures ($4000 - 4700$~K) for a $0.73 \pm 0.03 \ M_\odot$ main sequence star (see Figure \ref{sigGemHR}).  We obtain a range of luminosities ($0.11 - 0.15 \ L_\odot$) and radii ($0.70 - 0.59 \ R_\odot$).  We note that our analysis predicts a primary-to-secondary Johnson $V$-band flux ratio of $290$ assuming $T_\mathrm{eff,2} = 4500$~K), which is not in agreement with the flux ratio given by the spectroscopic 519~nm light ratio ($\sim70$ primary-to-secondary; see Appendix A for details on this measurement).  In order for our flux ratios to be in agreement, the secondary star would have $T_\mathrm{eff,2}=6400$~K, which is not consistent with the spectroscopic observations, nor with a main-sequence star given the location on the H-R diagram.  We cannot rule out the effect of starspots on the discrepant flux ratios as these were not accounted for when interferometrically detecting the companion and the spot features present during the interferometric and spectroscopic observations differ as evidenced in the APT light curve.   Additionally, \citet{pra02} and \citet{leh13} also reported discrepancies between TODCOR-reported flux ratios and their expected values.  Therefore, we use only the $H$-band flux ratio.  

We plot the location of the components of $\sigma$~Gem on an H-R diagram, as well as the corresponding evolutionary tracks.  We use Dartmouth stellar evolution tracks \citep[Fe/H $=0.0$, $\alpha$/Fe $=0.0$, PHOENIX-based models;][]{dot08} for the interpolated model masses ($M_{1,\mathrm{model}} = 1.28 \pm 0.07 \ M_\odot$,  $M_{2,\mathrm{model}} = 0.73 \pm 0.03 \ M_\odot$).  Our primary falls nearly on the $1.28 \ M_\odot$ evolutionary track with an estimated temperature of $4530 \pm 60$ K \citep{gle79,poe85,sta94,one96,kov01,mas08}.  The range of locations for the secondary on the H-R diagram passes through the main sequence for a star of $0.73 \ M_\odot$.  We find an age of the system of $5 \mp1$~Gyr.
Based upon the masses and age of the stars, we suggest that the primary star is an evolved late F-type star that is now a K giant.  The secondary star is a main-sequence early K star.

%%%%%%%%%%%%%%%%%%%%%%%%%%%%%%%%%%%%%%%%%%%%%%%%%%%%%%%%%%%%%%%%%%%%

\section{Ellipsoidal Variations and Gravity Darkening}

\citet{hen95} and \citet{kaj14} previously published subsets of the APT light curve data 
for starspot modeling and measuring differential rotation.  Both studies emphasized the presence of active longitudes on opposite sides of $\sigma$~Gem to explain the quasi-sinusoidal variation appearing at half of the orbital period.  

\begin{figure}
\hspace{-0.5cm}
\includegraphics[angle=90,scale=.38]{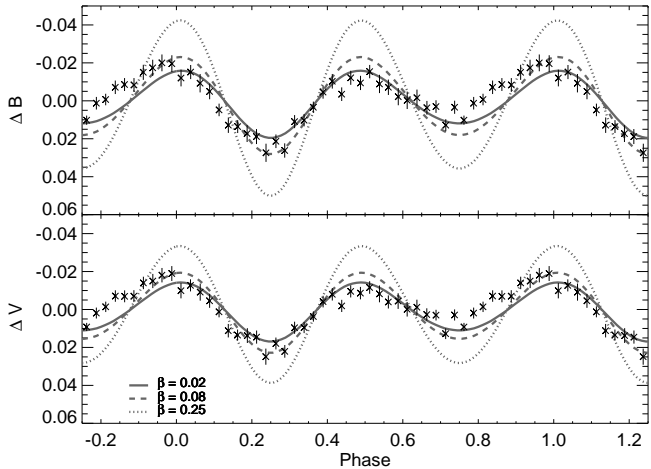}
\caption{Differential folded and binned light curves of $\sigma$~Gem for $B$ and $V$ magnitudes plotted with error bars from the binning.  Each data point is an average of data points spanning $0.025$ in phase from the complete light curve folded on the orbital period.  The quasi-sinusoidal signature of the averaged light curve is due to ellipsoidal variations caused by the primary star partially filling its Roche lobe potential.  The lines represents the ELC models for ellipsoidal variations with the gravity darkening coefficient $\beta = 0.02,0.08,$ and $0.25$, where $\beta = 0.02\pm0.02$ is the best fit to the binned and averaged light curves.  
}
\label{sigGemfit}
\end{figure}

We removed long-term trends, folded the APT photometry over the orbital period ($P_\mathrm{orb} = 19.6027$~days), and binned the data ($0.025$ in phase).  The resultant Johnson $B$ and $V$ light curves are presented in Figure \ref{sigGemfit}.  The quasi-sinusoidal trend observed in the averaged light curves suggests the possibility of ellipsoidal variations due to distortions of the primary star partially filling its Roche lobe potential.  With a Roche lobe radius of $16. 5 \ R_\odot$, we obtain $R_1/R_\mathrm{L} = 0.61$ \citep{egg83}.  

We used the light-curve-fitting software package Eclipsing Light Curve \citep[ELC;][]{oro00} to model the ellipsoidal variations using our orbital parameters with no free parameters \citep[gravity darkening assumed to be $\beta = 0.08$;][see Figure \ref{sigGemfit}]{luc67}.  The characteristics of the ellipsoidal variations with this model as compared to the light curve of $\sigma$~Gem indicate that the long-term signature likely is indeed due to ellipsoidal variations, in contrast to previous suggestions that the periodicity at $P_\mathrm{orb}/2$ is due to active longitudes aligned with the orbit \citep[e.g.,][]{hen95,jet96,ber98,kaj14,kov14}.  We note that rotation periods derived from the analysis of the light curve \citep[e.g.,][]{kaj14} suggest the star is rotating slightly faster than the orbital period, further supporting our identification of ellipsoidal variations in $\sigma$~Gem.  It should be noted that removing the effect of ellipsoidal variations from the light curve does not eliminate all starspot signatures (See Appendix B).

The ELC model fit of ellipsoidal variations can be improved to better match our data.  We modeled the system again with no free parameters except for the gravity darkening coefficient, $\beta$ for $T_\mathrm{eff} \propto g^\beta$ \citep{von24}, as \citet{esp12} recently suggested $\beta \sim 0.21$ for convective stars, substantially higher than the canonical $\beta \sim 0.08$ \citep{luc67} value assumed in our fixed-parameter fit.  Although our average light curve is still contaminated by some residual spot modulation, we find that $\beta = 0.02 \pm 0.02$ with error bars determined by bootstrapping over observing seasons of the 27 years of observation in the APT light curve.  This value strongly rules out $\beta > 0.1$ for this system (see Figure \ref{sigGemfit}).

%%%%%%%%%%%%%%%%%%%%%%%%%%%%%%%%%%%%%%%%%%%%%%%%%%%%%%%%%%%%%%%%%%%%

\section{Conclusions}

In this work, we have made the first visual detections of the secondary star of $\sigma$~Gem using interferometric and spectroscopic observations.  We establish the first visual orbit by combining the interferometric detections with radial velocity data.  The determination of orbital parameters has allowed for model-independent mass determinations ($M_1 = 1.28 \pm 0.07 \ M_\odot, M_2 = 0.73 \pm 0.03 \ M_\odot$).

Folded and binned photometric data have shown evidence of ellipsoidal variations, gravitational distortions of the primary star caused by the close companion.  The light curve is comparable to light curve models created only from stellar and orbital parameters (assuming no starspots).  Although the ellipsoidal variations are only a small effect, the primary star of $\sigma$~Gem is not spherical, partially filling its Roche lobe potential and having a surface temperature gradient.  Our establishment of ellipsoidal variations offers a compelling alternative explanation to the previously purported detections of active longitudes, starspots on either side of the primary star \citep{hen95,jet96,ber98,kaj14,kov14}.

Our new orbital elements along with the folded light curve also allow for measurements of gravity darkening.  We find that $\beta = 0.02\pm0.02$, a value of gravity darkening lower than suggested by theory \citep{luc67,esp11,esp12}.  

In this paper, we have demonstrated that precision interferometry at CHARA is now capable of detecting the faint main-sequence companions of bright RS CVn primary stars.  We are currently processing new data for other close, bright RS CVn systems and will be publishing these results in a series of follow-up papers.

\section*{Acknowledgements}

We thank  H.\ A.\ McAlister, J.\ A.\ Orosz, M.\ Reynolds, and M.\ Rieutord for their helpful comments and discussions.  The authors also thank Lou Boyd of Fairborn Observatory for the decades of crucial support he has given to our photometric program.  The interferometric data in this paper were obtained at the CHARA Array, funded by the National Science Foundation through NSF grants AST-0908253 and AST-1211129, and by Georgia State University through the College of Arts and Sciences.  The MIRC instrument at the CHARA Array was funded by the University of Michigan.  The photometric data were supported by NASA, NSF, Tennessee State University, and the State of Tennessee through its Centers of Excellence program.
R.M.R.\ would like to acknowledge support from the NASA Harriet G.\ Jenkins Pre-Doctoral Fellowship and a Rackham Graduate Student Research Grant from the University of Michigan.  
J.D.M.\ and R.M.R.\ acknowledge support of NSF grant AST-1108963.
A.N.A.\ and J.D.M.\ acknowledge the support of NSF AST-1311698.  
S.K.\ acknowledges support from an STFC Rutherford Fellowship (ST/J004030/1).
This research has made use of the SIMBAD database, operated at CDS, Strasbourg, France and the Jean-Marie Mariotti Center \texttt{SearchCal} service\footnote{Available at http://www.jmmc.fr/searchcal}
co-developped by FIZEAU and LAOG/IPAG, and of CDS Astronomical Databases SIMBAD and VIZIER\footnote{Available at http://cdsweb.u-strasbg.fr/}.

\appendix
\section{A. CfA Radial Velocities}

The CfA radial velocity data were obtained using two identical CfA Digital Speedometers
\citep{lat92} on two different telescopes: the $1.5$-m Wyeth
Reflector at the Oak Ridge Observatory located in the town of Harvard,
Massachusetts, and the $1.5$-m Tillinghast Reflector at the Fred
Lawrence Whipple Observatory (FLWO) on Mount Hopkins, Arizona.
In the first set of radial velocity measurements, altogether $78$ observations were obtained; $32$ with the Tillinghast
Reflector and $46$ with the Wyeth Reflector. Those radial velocities
have all been published \citep{mas08}, so the details of the
procedures and reductions will not be repeated here, except to provide
some overall characteristics of the data.  Forty of the observations
were obtained over a span of less than an hour using the Wyeth
Reflector.  They have been averaged to give one data point for the
analysis in this paper.  However, those forty observations provide an
opportunity to evaluate the precision of the individual velocity
measurements from the CfA Digital Speedometers for an RS CVn primary
with line broadening corresponding to a rotational velocity of $25$~
\kms.  The standard deviation of a single velocity from the average of
all $40$ is $0.40$~\kms.  This compares favorably with the value of $0.45$~
\kms reported by \citet{mas08} for the RMS velocity
residuals from their orbital solution using all $78$ velocities.

To avoid confusion, the velocities determined with the
CfA Digital Speedometers have always been published on the native
system of the instruments.  For the analysis in this paper we added
$0.14$~\kms to the published velocities, to put them on an absolute
system defined by the IAU Radial-Velocity Standard Stars \citep[][note
  that the sign of the correction was given as minus by mistake in
  that paper]{ste99}.

A new set of velocities for both components of the
$\sigma$~Gem system was obtained with the Tillinghast Reflector Echelle
Spectrograph \citep[TRES; ][]{fur08} at FLWO during the period 2012
October 1 to 2015 January 9.  TRES is a modern fiber-fed CCD echelle
spectrograph with resolution of $6.7$~\kms, very similar to that of the
CfA Digital Speedometers.  However, the free spectral range of the
echelle order centered at $519$~nm near the Mg~b features is $10$~nm,
compared to $4.5$~nm for the CfA Digital Speedometers.  For both
instruments this is the wavelength window used for the determination
of absolute velocities and for TODCOR analyses.  Furthermore the
signal-to-noise ratio per resolution element (SNRe) achievable with
the TRES CCD detector is much higher than was possible with the
intensified photon-counting detectors used in the CfA Digital
Speedometers.  The SNRe values for the old CfA spectra ranged from $30$
to $100$, while the typical value for the new TRES observations is $500$.

Fourteen strong TRES spectra were obtained of $\sigma$ Gem over a period
$30$ nights in 2014 December and 2015 January with the goal of detecting
the lines of the secondary and deriving a double-lined
spectroscopic orbit for the first time.  Fortunately, two earlier
observations from 2012 October were available in the TRES archive,
which provided a two-year baseline for determining a more accurate
orbital period.  All $16$ TRES spectra were analyzed using TODCOR
\citep{zuc94} as implemented at CfA by G.\ Torres, and using the
CfA library of synthetic spectra to choose the optimum templates.
Only one of the observations failed to give a reliable velocity for
the secondary, due to close blending of the lines from the two stars.

The light of the primary dominates the composite spectrum of $\sigma$~Gem,
so it was straightforward to choose the template that gave the highest
value for the average peak of the one-dimensional correlations; the
parameters for the best template were $T_{\mathrm{eff},1} = 4500$~K, $\log g_1 = 2.5$
(cgs), $v \sin i_1 = 25$~\kms, and solar metallicity.  Because the
secondary is so much fainter, its template parameters were only weakly
constrained by the two-dimensional correlations.  We therefore adopted
the mass derived in this paper to guide the choice of secondary
template parameters, $T_{\mathrm{eff},2} = 4250$~K, $\log g_2 = 4.5$ (cgs), $v \sin i_2 =
2$~\kms, and solar metallicity; the rotational velocity was selected
under the assumption of tidal synchronization of the secondary spin
with the orbital period.  We also tried TODCOR solutions using the
nearest neighbors from our library for the secondary template, and
found that the results were not sensitive to the secondary template
parameters.

TODCOR has a mode in which the light ratio between the secondary and
the primary in the observed wavelength window can be treated as a free
parameter and thus can be determined from the spectra.  We selected $10$
observations where the velocities of the primary and secondary were
well separated and used these to determine a light ratio at $519$~nm of
$0.0139 \pm 0.0025$ (standard deviation of the values from the mean).
If the errors are well behaved in this analysis, the uncertainty of
the mean light ratio from TODCOR could be formally better by about a
factor of 3, or about $6 \%$.  For the final TODCOR analysis we
fixed the light ratio to $0.0139$ for all the observations.  The TODCOR
velocities and times of observation are reported in Table \ref{RVCfA} on the
same zero point as the native CfA Digital Speedometer system.  For the
analysis in this paper, $0.14$~\kms was added to the velocities in
Table \ref{RVCfA} to put them on the IAU system.

The orbital parameters are reported in Table \ref{RVonlyorb} for two different
double-lined spectroscopic orbits based on just the TODCOR velocities
from the $16$ TRES spectra. For the first solution the eccentricity was
allowed to be a free parameter.  Since the derived eccentricity was
not significant, $e = 0.0018 \pm 0.0029$, we also derived a solution
for a circular orbit.  Note that the mass ratio is well constrained;
$q = 0.582 \pm 0.016$ ($2.7 \%$).  Table \ref{RVonlyorb} also contains the orbit reported by \citet{mas08} and the orbit fit with only the data from the AST/TSU data set.  Like the results from the TRES
spectra, the eccentricity from the AST radial velocities,
$e = 0.002 \pm 0.003$, is consistent with a circular orbit.  Note that these are the entire data sets collected at these facilities and contain radial velocity measurements that were obtained when starspots were present on the stellar surface.  
\begin{turnpage}
\begin{deluxetable}{l c c c c c}
\tablecaption{$\sigma$~Gem Orbital Parameters from Radial Velocity Curves$^a$}
\tablehead{
\colhead{Parameter}         &
\colhead{TRES/TODCOR}          &
\colhead{TRES/TODCOR}      &
\colhead{\citet{mas08}} &
\colhead{AST/TSU} 
&
\colhead{AST/TSU}  \\
& \colhead{$e$ free} & \colhead{$e$ fixed} & \colhead{$e$ free} & \colhead{$e$ free} & \colhead{$e$ fixed}
}
\startdata
orbital period,             $P_\mathrm{orb}$ (days)         &$ 19.6059 \pm 0.0020 $&$ 19.6065 \pm 0.0018 $ & $19.60437 \pm0.00053$ & $19.6041 \pm 0.0002 $& $19.6041\pm0.0002$ \\
center of mass velocity,    $\gamma$ (\kms)    &$ 43.554 \pm 0.077   $&$ 43.553 \pm 0.074   $ & $43.043\pm0.066$ &$43.25 \pm 0.07$& $43.33\pm0.06$\\
semi-amplitude, primary,    $K_\mathrm{1}$ (\kms) &$ 35.19 \pm 0.10     $&$ 35.192 \pm 0.098   $ & $34.776\pm0.100$ &$34.60 \pm 0.08$& $34.60\pm0.08$ \\
semi-amplitude, secondary,  $K_\mathrm{2}$ (\kms) &$ 60.5 \pm 1.6       $&$ 60.5 \pm 1.5       $ &  & &  \\
eccentricity,               $e$                &$ 0.0018 \pm 0.0029  $&$  0           $ & $0.0143\pm0.0026$ &$0.002 \pm 0.003 $& 0\\
time of periastron passage, $T$ (HJD)          &$ 2456899.0 \pm 5.9  $& & $2453507.96\pm0.71$ & $2456985.3 \pm 3.7$ &  \\
longitude of periastron,         $\omega_\mathrm{1}$ ($^\circ$) &$38. \pm 109     $& & $46\pm13$ & $181 \pm 68$&  \\
time of maximum velocity    $T_0$ (HJD)  & &$ 2456916.571 \pm 0.012 $ & & &  $2453563.81\pm0.02$\\
$M_\mathrm{1} \sin^3 i$  ($M_\odot$)                  &$ 1.126 \pm 0.068    $&$ 1.126 \pm 0.065    $ & & &\\
$M_\mathrm{2} \sin^3 i$  ($M_\odot$)                  &$ 0.655 \pm 0.022    $&$ 0.655 \pm 0.021    $ & & &\\
$a_\mathrm{1} \sin i$ ($10^6$ km)                 &$ 9.487 \pm 0.027    $&$ 9.488 \pm 0.027    $ & $9.374\pm0.024$ & $9.33\pm0.02$ & $9.33\pm0.02$ \\
$a_\mathrm{2} \sin i$ ($10^6$ km)                 &$ 16.31 \pm 0.43     $&$ 16.32 \pm 0.42     $ & & & \\
$a \sin i$ ($R_\odot$)                             &$ 37.09 \pm 0.62     $&$ 37.09 \pm 0.60     $ & & &\\
mass ratio, $q = M_2/M_1$                                &$ 0.582 \pm 0.016    $&$ 0.582 \pm 0.15     $ & & &\\
mass function, $f(M) = (M_2 \sin i)^3/(M_1 + M_2)^2$ & & & $0.0854 \pm 0.00066$ & $0.0841 \pm 0.0006$ & $0.0841\pm0.0006$ \\
RMS velocity residuals, $\sigma_\mathrm{1}$ (\kms) &$ 0.29              $&$ 0.28               $ & $0.45$ & $0.30$ & $0.30$ \\
RMS velocity residuals, $\sigma_\mathrm{2}$ (\kms) &$ 4.8               $&$ 4.6                $ & & &\\
light ratio at $519$~nm                           &$ 0.0139            $&$ 0.0139             $ & & & 
\enddata
\tablecomments{$^a$These radial velocity data sets are the complete sets and include data contaminated by starspots.
}
\label{RVonlyorb}
\end{deluxetable}
\end{turnpage}

Our double-lined orbital solution is among the most extreme that we
have derived using the Mg~b region in terms of the light ratio.  Indeed, attempts to get good solutions from
the neighboring echelle orders on either side of the Mg~b order were
unsatisfactory.  Therefore we decided it would be prudent to test the
reliability of our TODCOR analysis by creating sets of simulated
observations using our library of synthetic spectra.  For the time of
each observed spectrum we simulated that observation by shifting the
two synthetic template spectra by the velocities from the orbital
solution and coadding after scaling by the light ratio found by TODCOR
from the real observations.  We found that a TODCOR analysis of the
simulated observations reproduced the mass ratio and light ratio of
the real data well within the estimated errors, and that the mass
ratio was not sensitive to the value chosen for the light ratio.  Thus
we have no evidence for a systematic error in our determination of a
light ratio of $0.0139$ at $519$~nm.  This is quite different from the
light ratio reported for the interferometric observations of $0.0038$ at
$H$-band.

\section{B.  Difference Light Curves}

\begin{figure}
\hspace{3.5cm}
\includegraphics[angle=90,scale=.45]{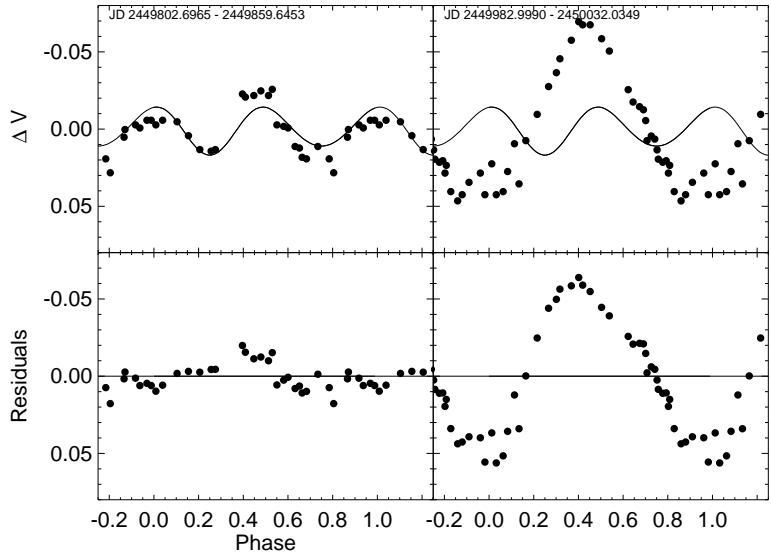}
\caption{Average-subtracted, differential light curves of $\sigma$~Gem for Johnson $V$ magnitudes plotted for JD $2449802.6965 - 2449859.6453$ (left; Segment 8, Set 45 \citep{kaj14}) and JD $2449982.9990 - 2450032.0349$ (right; Segment 9, Set 1 \citep{kaj14}).  The top panel contains a plot of the APT data sets (circles) and the model ellipsoidal variations created with ELC for the orbital parameters of $\sigma$~Gem and best-fit gravitational darkening coefficient $\beta = 0.02$ (solid line).  The bottom panel contains the residuals of the APT light curve with the ellipsoidal variation signature removed (circles).  
}
\label{EVremove}
\end{figure}

In order to better justify our conclusion that ellipsoidal variations can explain previous claims of ``active longitudes'' on $\sigma$~Gem, we have re-plotted some photometry from \citet{kaj14} in Figure \ref{EVremove} along with our prediction of the expected ellipsoidal variation component using the ELC software and system parameters from Table \ref{sigGemparam} using gravity darkening parameter $\beta=0.02$.  In Figure \ref{EVremove}, we include data from two epochs, one showing very little overall variability and one showing high variability. In the first epoch (``Segment 8, Set 45'' of \citet{kaj14}), the photometric data showed clearly a double-peaked light curve when phased with the orbital period, previously interpreted as due to active longitudes \citep[see][]{kaj14}. Here, we now see by removing the expected ellipsoidal variation, the signature of two spots on opposite sides of the star (the basis for the active longitudes claims) nearly completely disappears (see Figure \ref{EVremove}).  The second epoch (``Segment 9, Set 1'' of \citet{kaj14}) is dominated by one spot and the ellipsoidal variations are not discernible.  Nonetheless, future starspot modelers should account for the underlying ellipsoidal variations before performing detailed light curve analysis or surface brightness inversions. We have included our calculation of the expected ellipsoidal variations for Johnson $UBVRIJHK$ in Table \ref{EVfilters} to assist future workers--a full re-analysis of the star spot properties is beyond the scope of this paper.

\begin{deluxetable}{l c c c c c c c c}
\tabletypesize{\scriptsize}
\tablecaption{Johnson Photometric Ellipsoidal Variations Models of $\sigma$~Gem}
\tablewidth{0pt}
\tablehead{
\colhead{Phase} & \colhead{$U$} & \colhead{$B$} & \colhead{$V$} & \colhead{$R$} & \colhead{$I$} & \colhead{$J$} & \colhead{$H$} & \colhead{$K$}
}
\startdata
     0.000 &     -0.018 &     -0.016 &     -0.014 &     -0.013 &     -0.012 &     -0.011 &     -0.011 &     -0.010 \\
     0.011 &     -0.019 &     -0.016 &     -0.014 &     -0.013 &     -0.012 &     -0.011 &     -0.011 &     -0.010 \\
     0.022 &     -0.019 &     -0.016 &     -0.014 &     -0.013 &     -0.012 &     -0.011 &     -0.010 &     -0.010 \\
     0.033 &     -0.018 &     -0.015 &     -0.014 &     -0.012 &     -0.011 &     -0.011 &     -0.010 &     -0.009 \\
     0.044 &     -0.017 &     -0.015 &     -0.013 &     -0.012 &     -0.011 &     -0.010 &     -0.009 &     -0.009  
\enddata
\tablecomments{Table \ref{EVfilters} is published in its entirety in the electronic edition.  Orbital phase is based upon our orbital parameters listed in Table \ref{sigGemparam}.  Notably, $T_0 = 2453583.98$ (HJD), $P_\mathrm{orb} = 19.6027$ days. } 
\label{EVfilters}
\end{deluxetable}

%%%%%%%%%%%%%%%%%%%%%%%%%%%%%%%%%%%%%%%%%%%%%%%%%%%%%%%%%%%%%%%%%%%%

\end{document}